\documentclass[runningheads]{llncs}

 \usepackage{soul}
 \usepackage{amsmath}
 \usepackage{algorithm}
\usepackage{algorithmic}
\usepackage{longtable}
\usepackage{multicol}
\usepackage{comment}
\usepackage{graphicx}
\graphicspath{ {./Images/} }

\usepackage[subtle]{savetrees}

\AtBeginDocument{%
  \providecommand\BibTeX{{%
    \normalfont B\kern-0.5em{\scshape i\kern-0.25em b}\kern-0.8em\TeX}}}

\makeatletter
\renewcommand\section{\@startsection{section}{1}{\z@}%
                       {-8\p@ \@plus -4\p@ \@minus -4\p@}%
                       {6\p@ \@plus 4\p@ \@minus 4\p@}%
                       {\normalfont\large\bfseries\boldmath
                        \rightskip=\z@ \@plus 8em\pretolerance=10000 }}
\renewcommand\subsection{\@startsection{subsection}{2}{\z@}%
                       {-8\p@ \@plus -4\p@ \@minus -4\p@}%
                       {6\p@ \@plus 4\p@ \@minus 4\p@}%
                       {\normalfont\normalsize\bfseries\boldmath
                        \rightskip=\z@ \@plus 8em\pretolerance=10000 }}
\renewcommand\subsubsection{\@startsection{subsubsection}{3}{\z@}%
                       {-4\p@ \@plus -4\p@ \@minus -4\p@}%
                       {-1.5em \@plus -0.22em \@minus -0.1em}%
                       {\normalfont\normalsize\bfseries\boldmath}}
\makeatother

\begin{document}

\title{QR-SACP: Quantitative Risk-based Situational Awareness Calculation and Projection through Threat Information Sharing}

\author{Mahdieh Safarzadehvahed \inst{1} \and Farzaneh Abazari \inst{1} \and
Afsaneh Madani\inst{1}  
\and 
 \\ Fatemeh Shabani \inst{1}
}

\authorrunning{M. Safarzadehvahed et al.}

\institute{Iran Telecommunication Research Center, Tehran, Iran \\
\email{\{m.safarzadeh, f.abazari, madani\}@itrc.ac.ir}, \\ 
\email{\{fateme.shabani\}@modares.ac.ir}}




\maketitle

\begin{abstract}
When a threat is observed, one of the most important challenges is to choose the most appropriate and adequate timely decisions in response to the current and near future situation in order to have the least consequences and costs. Making the appropriate and sufficient decisions requires knowing what situations the threat has engendered or may engender. In this paper, we propose a quantitative risk-based method called QR-SACP to calculate and project situational awareness in a network based on threat information sharing. In this method, we investigate a threat from different aspects and evaluate the threat's effects through dependency weight among a network's services. We calculate the definite effect of a threat on a service and the cascading propagation of the threat’s definite effect on other dependent services to that service. In addition, we project the probability of a threat propagation or recurrence of the threat in other network services in three ways: procedurally, network connections and similar infrastructure or services. Experimental results demonstrate that the QR-SACP method can calculate and project definite and probable threats’ effects across the entire network and reveal more details about the threat's current and near future situations.
 
\end{abstract}

%



\section{Introduction}

When a threat is observed, one of the most important challenges is to choose the most appropriate and adequate timely decisions, so-called decision-making in response to the situation that the threat has caused or may cause in the future. Decision-making in response to the threat has different consequences and costs. What decision to make and when to deal with it in the threat cycle directly impacts the cost and the damage it may cause. Making the appropriate and adequate decision requires knowing what the threat is, what impacts it has, and what situation the threat has engendered or may engender. One of the concepts that can help is situational awareness (SA). In the field of SA, various studies~\cite{alavizadeh2021survey,alavizadeh2020cyber,franke2014cyber,jajodia2009cyber,pahi2017analysis,poyhonen2020cyber,skopik2012designing,solutionsoverview,zhang2018network,Zhang2021Network} have been done on different parts of SA, including definition, architecture, modeling, uncertainty and risk management, projection and calculation. Despite having the current and the future situations that a threat has posed or may pose, there is still no proper answer about which threat should be investigated first, how SA should be quantified and how probable near future threats' effects should be considered in calculating and projecting the SA.

In this paper, we present a novel algorithm for calculating and projecting SA that by knowing the past and receiving a threat, calculates the SA to depict the present and projects the consequences of the threat to predict the future through a risk-based approach. The obtained quantitative SA values can be used to select a high-priority threat to investigate.

One of the most important factors for decision-making is to get an accurate and comprehensive view of the current situation of the entire network due to a threat. We present an algorithm that perceives and calculates the impacts of a threat across the entire network using threat information sharing, organizational historical information about threats and international threats databases. It is also necessary to achieve a comprehensive and integrated view of the under attack network and its connected networks to make a comprehensive decision. This requires understanding what has happened or will happen due to the threat. To achieve this goal, we investigate the threat from different dimensions and calculate and project its impacts.
Making inappropriate and insufficient decisions can lead to the continuation of the threat and its impacts. For this reason, it is necessary to project the near future situation in addition to understanding the current situation that arises when the threat occurs. In contrast to earlier algorithms and methods, we project a risk-based near future situation with more details in the proposed algorithm. 

The number of seen threats in a network is very high. Hence, we need a way to help us select a threat with more priority. Most studies display SA qualitatively in the form of green, blue, yellow, orange, and red. In this case, a large number of threats will be categorized into one color group. Therefore, we cannot choose a threat with a higher priority among them to investigate. In addition, two threats may have the same color but differ in properties. As a result, providing a color is not enough to select high-priority threats. Hence, we calculate and display the SA quantitatively by considering more details.
Our main contributions are as follows. (1) We propose an algorithm that investigates each threat from different dimensions and uses service dependency among network's services to calculate the definite effect of the threat on the service and the cascading propagation of the threat’s definite effect to other dependent services with more details. (2) We propose an algorithm for projecting the probable impacts of the threat on the network and predicting a risk-based near future network situation. The algorithm projects the probability of propagation or recurrence of the threat in other dependant network services. (3) We propose a method to map a network's situational awareness to a four elements vector. Each part of it reveals different, various, definite and probable effects of a threat on a network. These four parts provide a comprehensive view of the network's situational awareness.

The remainder of the paper is as follows. In Section ~\ref{sec:Relwork}, we review some related works and discuss their limitations in calculating and projecting SA. In Section ~\ref{sec:ASSmdl}, we present assumptions and concepts to calculate SA and model a network and a threat. In Section ~\ref{sec:QR-SAC}, we propose QR-SACP, a novel technique that uses service dependencies and probability of propagation or recurrence of the threat to calculate and project SA, and then we evaluate the effectiveness of QR-SACP by using various threats in section ~\ref{sec:result}. Finally, we draw conclusions in Section ~\ref{sec:CONCLUSION}.

\section{Related Work}\label{sec:Relwork}

Some studies~\cite{ahmad2021can,alavizadeh2021survey,franke2014cyber,jajodia2009cyber} have done a comprehensive literature review that provide information about different aspects of situational awareness, including SA models, frameworks, architecture and uncertainty management and attack prediction.
Since this paper presents a method to calculate and project SA quantitatively, related works have been selected in such a way that they have been done in the field of SA calculation and projection.

Zhang et al.~\cite{Zhang2021Network} present a network SA model based on threat intelligence to conduct situational perception and discover internal threats. They collect situation elements of network asset status, risk status, and log warnings. However, they do not specify what details and parameters these inputs contain. They filter and clean collected data and correlate them with external threat intelligence to find internal threats. They use game theory to quantify the current network security situation of the system and evaluate the network security status. They also use Nash equilibrium to predict attacks behavior. One of the most important parts of this model is to calculate the situational awareness based on attacker and defender strategies, but they do not introduce these strategies. In this study, situational awareness is equivalent to the difference between an attacker and defender utility. However, this approach does not provide how the attacker and defender utility is calculated, and they discard dependency among systems in a network.

Alavizadeh et al.~\cite{alavizadeh2020cyber} introduce a framework to select a response strategy in order to defend against possible attacks. They consider two defense strategies: Virtual Machine Live Migration (VM-LM) and Patching. To choose a defense strategy, they calculate risk values. For this purpose, they propose three security metrics: 1) risk of exploiting a VM, 2) security Return on the Attack (RoA) and 3) Mean of Attack Path Length (MAPL). They select the defense strategy based on the risk of exploiting a VM security metric. They suppose SA calculation is equivalent to risk calculation by considering vulnerabilities. They ignore attacks' and incidents' effects on VMs to select the defense strategy. The risk of spreading a threat is not only due to a vulnerability on the victim system or network, but it can also be due to obtained privileges that an attacker gains. After gaining privilege, an attacker no longer needs to exploit a vulnerability to access another system because he can continue his objectives by performing authorized actions with the obtained privileges. Moreover, the value of assets is not considered in the calculation of situational awareness. 

Rongrong et al.~\cite{rongrongframework} propose a framework to evaluate a network security situation through three dimensions: threat, vulnerability and stability. They calculate the average value for each dimension and merge these three dimensions' results to measure the overall network security situation. They consider the successful probability of attacks and their severity to assess threats. To evaluate network vulnerability situation, they consider vulnerability vendor name, product name, type, severity and the duration between the data a vulnerability is disclosed and a patch is released for it. They consider TCP, UDP and ICMP input and output traffic to assess stability. They do not consider dependencies among systems or services in threat assessment. Therefore, they cannot calculate an attack propagation impact. Furthermore, the average of items is calculated instead of adding the calculated value for each item in each dimension with the other items in that dimension.

Kou et al.~\cite{kou2019research} present a method to evaluate a network security situation based on attack intention. The method recognizes the attack intention and attack stages, calculates the network SA and predicts the next attack stage based on achieved attack stages. This method determines attack paths, then calculates the SA for each path by multiplying the probability of the attack stage with the destructiveness of the attack and the weight of the node in which the attack occurred. Finally, it sums the SA of each attack path together to obtain the SA of the whole network. The research method of this paper depends on the known attack patterns. Therefore, it cannot calculate the network SA for unknown attacks. They do not consider systems' defense measures in the network to calculate SA. In addition, they have considered the existence of a vulnerability as the only reason for spreading the threat while the procedural relationship and the existence of similar assets can cause threat recurrence. 

Zhang et al.\cite{zhang2018network} present a framework for assessing network security situational awareness in cloud computing through stochastic games. They predict the attack behavior using a fuzzy optimization method and Nash equilibrium. This method has been provided for use in the cloud computing environment. In addition, to determine situational awareness, this method considers threatening failure on a host and does not consider its propagation in the network due to the dependency among hosts and services they provide. It also does not consider the possibility of spreading threats in the network. 

Xiao-Lu et al.~\cite{han2018research} propose a model and methodology to assess big data security situation. This model includes an index system and a fuzzy comprehensive evaluation algorithm to assess big data security situations. They consider two index levels. The first index level reflects the big data security situation from four dimensions, including the damage degree of harmful procedures, the damage degree of information destruction, the degree of menace and the damage degree of attacks. The second index level is designed according to the first level but with finer granularity. They use the fuzzy comprehensive evaluation algorithm to assess big data security situations. First, they identify big data security important features. Then they establish measurable index factor sets. In the third step, they establish a measurement level. After that, they establish a fuzzy relation matrix. In the fifth step, they determine the weight vector of the index. Next, they calculate big data security situations comprehensive evaluation.

Marcus et al.~\cite{pendleton2016survey} propose measuring system-level security through a security metrics framework based on metrics of system vulnerabilities, defense mechanisms, threat severity and situations. To investigate the relationships among these four sub-metrics, they propose a hierarchical ontology with four sub-ontologies corresponding to the four sub-metrics. They calculate a network SA at time $t$ as a function of $V(t)$, $D(t)$ and $A(t)$ which are a function of vulnerabilities, defenses and attacks at time $t$, respectively.

\section{ASSUMPTIONS and MODELING}\label{sec:ASSmdl}
The proposed situational awareness computational model has been presented to use in an ICT-ISAC, the largest available network in each country, yet this model can be used in any network, too. In the following, wherever we mention the network, we mean the ICT-ISAC network. Since in such a network, each service may be provided by more than one organization, such as the Internet service provided by several ISPs, for better understanding and readability, here we display each service as a service-organization.

In this paper, we want to calculate SA quantitatively. 
Endsley~\cite{endsley1988design} defined SA as “the perception of the elements in the environment within a volume of time and space, the comprehension of their meaning and the projection of their status in the near future.” Hence, to calculate SA, we should understand the impacts of a threat on a network and the projection of risks that the threat may pose in the near future on the network. Each observed threat may have a definite effect on the network or put the network at risk in the near future taking into account the historical information of the monitoring network, international threats databases include NVD~\cite{nvd:2021}, CVE~\cite{cvedetails:2021} and CAPEC~\cite{capec:2021} and monitoring devices. 

To calculate SA, data must be collected from a set of monitoring devices and analyzes must be performed. In this paper, we focus on how SA is calculated. Our purpose is to monitor the situation of critical services although all services can be monitored. Therefore, we calculate the SA of threats that affect critical services. Therefore, we suppose the network’s assets are critical services that from now on, we call them services.
\subsection{Modeling Network}\label{subsec:Modeling Network}
\begin{figure}
\centering
\begin{minipage}{.5\textwidth}
  \centering
  \includegraphics[width=.4\linewidth]{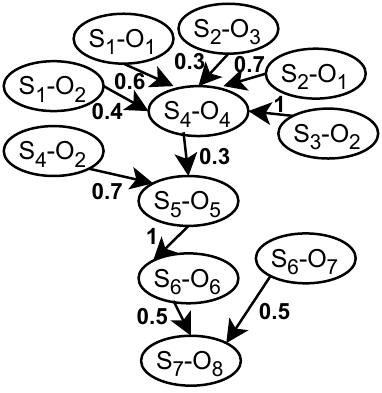}
  \caption{A Typical Service Dependency Graph in a Network}
  \label{fig:Fig1}
\end{minipage}%
\begin{minipage}{.5\textwidth}
 \centering
  \includegraphics[width=.4\linewidth]{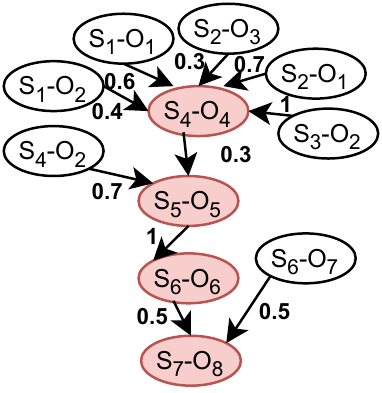}
  \caption{A threat effects on services using the cascade propagation}
  \label{fig:Fig4}
\end{minipage}
\end{figure}

Let $\mathcal{N}$ be a network of different organizations that provides services to users as shown in Figure ~\ref{fig:Fig1}. We represent each provided service by an organization by $S_i-O_j$ and call it service-organization $S_i-O_j$. Each organization can provide more than one service, and some organizations may rely on services provided by other organizations to offer some of their services. For example, the organization that provides the SIM card sales service is dependent on the person profile inquiry service from the Civil Registration Organization. We model network $\mathcal{N}$ by a weighted directed graph whose nodes represent the service-organizations in the network, whose directed edges represent the dependency relationships between them and the weight of each directed edge represents the value of the dependency relationship. Formally, a service dependency graph is a pair $\mathcal{G}=(S,D)$, where $S\subseteq\mathcal{S}$ is a finite set of nodes or service-organizations and $D\subseteq S\times S$ is a finite set of directed edges between the nodes of the graph. 
In this study, we represent each service by a $12$-tuple $(Sid, Oid, Crit, P_e, {Destsrv, Prt, Proto, Dir, W}, Confdemand$, $Integdemand$, $Avldemand)$ in which $Sid$ and $Oid$ are the identifiers of the service and the organization which provides the service, respectively. Also, $Crit$ is the criticality of the service in the network, $Destsrv$ is the service which depends on the service $Sid$, $P_e$ reflects the service's security controls to protect it against the threats, $Prt$ is port number of the link between $Sid$ and $Destsrv$, $Proto$ is the protocol of the link between $Sid$ and $Destsrv$, $Dir$ is the direction of the link between $Sid$ and $Destsrv$, $W$ is the weight of dependency between $Sid$ and $Destsrv$, $Confdemand$, $Integdemand$ and $Avldemand$ are the required value for Confidentiality, Integrity and Availability of service $Sid$ that the service should have to be secure, respectively. More formally, for any given service $s$, we write $Sid(s)$, $Oid(s)$, $Crit(s)$, $Destsrv(s)$, $P_e(s)$, $Prt(s)$, $Proto(s)$, $Dir(s)$, $W(s)$, $Confdemand(s)$, $Integdemand(s)$, $Avldemand(s)$ to denote its associated $id$, organization, destination service, defensive probability, port, protocol, link direction, dependency weight and demand Confidentiality, Integrity and Availability, respectively.
We also represent each organization by a $6$-tuple $(Oid, Crit, P_e, Confdemand, Integdemand, Avldemand)$ in which $Oid$ is the identifier of the organization, $Crit$ is the criticality of the organization, $P_e$ reflects the organization's security controls to protect it against the threats, $Confdemand$, $Integdemand$, $Avldemand$ are the required value for Confidentiality, Integrity and Availability of organization that the organization should have to be secure, respectively. More formally, for any given organization $o$, we write $Oid(o)$, $Crit(o)$, $P_e(o)$, $Confdemand(o)$, $Integdemand(o)$, $Avldemand(o)$ to denote its associated identifier, criticality, defensive probability and demand Confidentiality, Integrity and Availability, respectively.
\begin{example}
In Figure~\ref{fig:Fig1}, when the weight of dependency relationship between $S_6-O_6$ and $S_7-O_8$ is $0.5$ means organization $O_8$ is fifty percent dependent on service $S_6$ of organization $O_6$ to provide service $S_7$.
\end{example}
\subsection{Modeling Threat}
Various threats target networks. This diversity can be examined in terms of a threat's type, severity, complexity and effects that the threat has on a network. NIST defines a threat as follows~\cite{NISTSP800-30:2012}: "A cyber threat is any circumstance or event with the potential to adversely impact organizational operations (including mission, functions, image, or reputation), organizational assets, individuals, other organizations, or the Nation through an information system via unauthorized access, destruction, disclosure, modification of information, and/or denial of service.”
We divide each threat into the following three categories based on the above definition and the effects that threat has and the probable consequences that come with it:
\begin{itemize}
    \item A vulnerability is a threat that does not currently affect the network, but if it exists in an asset, it can have consequences in the future
    \item An attack is a threat that can have probable impacts on the network and can have consequences in the future
    \item An incident is a threat that has definite effects and may have consequences in the future. 
\end{itemize}
When a threat is observed, it is necessary to investigate the threat and its impacts to calculate and project SA and make the best decision to deal with it. We can investigate each threat from four dimensions, each threat 1) has some properties, 2) has some impacts, 3) occurs in an infrastructure and configuration and 4) can propagate through the network and infect other connected systems and networks. Hence, we should consider the mentioned dimensions to calculate and project SA. We introduced threat types earlier, and now we define four dimensions of a threat in the following. 
\subsubsection{Threat Properties}
Each threat has some properties which is defined by them. In this study, we represent each threat by a $15$-tuple $(Tid, Type, Vulid, Atkid,$ $Name, P_A, AConfimp, AIntgimp, AAvlimp,$ $Sid, SnsName, Cat, Prt,$ $Proto, CPEid)$ in which $Tid$, $Vulid$, $Atkid$, $Sid$ and $CPEid$ are the identifiers of the reported threat, vulnerability in CVE~\cite{cvedetails:2021}, attack in CAPEC~\cite{capec:2021}, service and asset on which threat is observed, respectively. Also, Type is the type of the threat which can be vulnerability, attack or incident, Name is the name of vulnerability, attack or incident, $P_A$ is the probability of successful occurrence of the threat, $AConfimp$, $AIntgimp$ and $AAvlimp$ are the announced impacts by the organization which threat has had on the service certainly, respectively, $SnsName$ is the name of a sensor that has detected the threat, $Cat$ is the category that the threat belongs to, $Prt$ and $Proto$ are port number and protocol which threat uses them. More formally, for any given threat $t$, we write $Tid(t)$,$Type(t)$, $Vulid(t)$, $Atkid(t)$, $Name(t)$, $P_A(t)$, $AConfimp(t)$, $AIntgimp(t)$, $AAvlimp(t)$, $Sid(t)$, $SnsName(t)$, $Cat(t)$, $Prt(t)$ ,$Pro-$ $to(t)$, $CPEid(t)$ to denote its associated Identifier, Type, Vulnerability, Attack, Name, successful occurrence probability, definite Confidentiality, Integrity and Availability impact, service, detector sensor name, category, port, protocol and asset,  respectively. We extract these properties from CVE~\cite{cvedetails:2021}, NVD~\cite{nvd:2021} and CAPEC~\cite{capec:2021}.
\subsubsection{Threat Occurrence Infrastructure and Configuration}
Each threat can occur in a service in an organization, but services can be equipped with security controls, for example, preventive security tools, secure configurations, best practices and the like. Hence, the threat may not occur successfully because of security controls. As a result, security controls should be considered besides threat properties to calculate SA. For instance, when a threat targets port 80, but it is reported closed in security controls, the threat is unsuccessful. Therefore, we should take threat occurrence infrastructure into account in the SA calculation.
As mentioned in Section ~\ref{subsec:Modeling Network} we use $P_e$ parameter to show each service or organization defensive probability against threats that reflects security controls that there are on them to protect them. 
\subsubsection{Threat Impacts}\label{subsec:Threat Impacts}
Although each threat has its own effects, it may have different effects in practice. For each threat, we can have two types of impacts:
\begin{itemize}
\item The potential impact, once a threat is observed in a service, it may have some impacts on the service’s Confidentiality, Integrity and Availability. The impacts that each threat has on a service potentially in case of the successful vulnerability exploitation is called the "Adjusted Impact" that we represent it by $CIAI$. We calculate Adjusted Impact by using Equation~\ref{Eq:Eq1}~\cite{doynikova2017cvss}. We obtain $Confimp(t)$, $Intgimp(t)$, $Avlimp(t)$ values from $Vulid(t)$ specifications. 
\item The affected impact, each threat may not have all of its impacts on a service because of applying security controls. We name impacts that each threat certainly has on a service the "Affected Impact" and represent it by $CIAAI$ for threat $t$. We represent these impacts for threat $t$ by $AConfimp(t)$, $AIntgimp(t)$ and $AAvlimp(t)$ which are announced by the organization as mentioned earlier. We calculate the Affected Impact through Equation~\ref{Eq:Eq1} by replacing $Confimp(t)$, $Intgimp(t)$, $Avlimp(t)$ values by $AConfimp(t)$, $AIntgimp(t)$ and $AAvlimp(t)$. 
\end{itemize}
\begin{equation}\label{Eq:Eq1}
CIAI(t)= min (10, 10.41 \times (1- (1- Confimp(t)) \times (1- Intgimp(t)) \times (1- Avlimp(t))))
\end{equation}
\subsubsection{Threat Propagation}\label{subsec:Threat Propagation}
When a threat occurs in an isolated service-organization, it has a different SA than in a service-organization that is connected to other service-organizations. Service-organizations may have similarities in infrastructure or have communications together which may lead to the propagation and repetition of a threat from one service-organization to other service-organizations associated with it. Therefore, the threat can also propagate to other service-organizations and affect them. Each threat may propagate in a network through three following methods:
\paragraph{Threat propagation procedurally}
Organizations in a network may exchange information together procedurally. This means they send data and information to each other via email, automation systems, sending USB flash dirves. If a threat occurs in an organization, the organization may transfer the threat to other organizations with which it communicates.
\paragraph{Threat propagation through network connections}\label{subsubsubsec:Threat propagation through network connections}
An organization in a network may connect to other organizations through network connections and provide services to them or benefit from their services. When a threat occurs in an organization, the threat may be transmitted from that organization to organizations connected with it through the network. In addition, an attacker may infiltrate an organization then infiltrate other organizations by exploiting the first organization’s connections with them.
\paragraph{Recurrence of a threat in other organizations due to similar infrastructure or services}
Each organization has a specific infrastructure, including hardware and software, or provides some services. If a threat occurs on hardware, software or a service in an organization, this threat may occur in other organizations with similar infrastructure or services.

\section{QUANTITATIVE RISK-BASED SITUATIONAL AWARENESS CALCULATION AND PROJECTION}\label{sec:QR-SAC}
\begin{figure}
    \centering
    \includegraphics[width=0.7\textwidth]{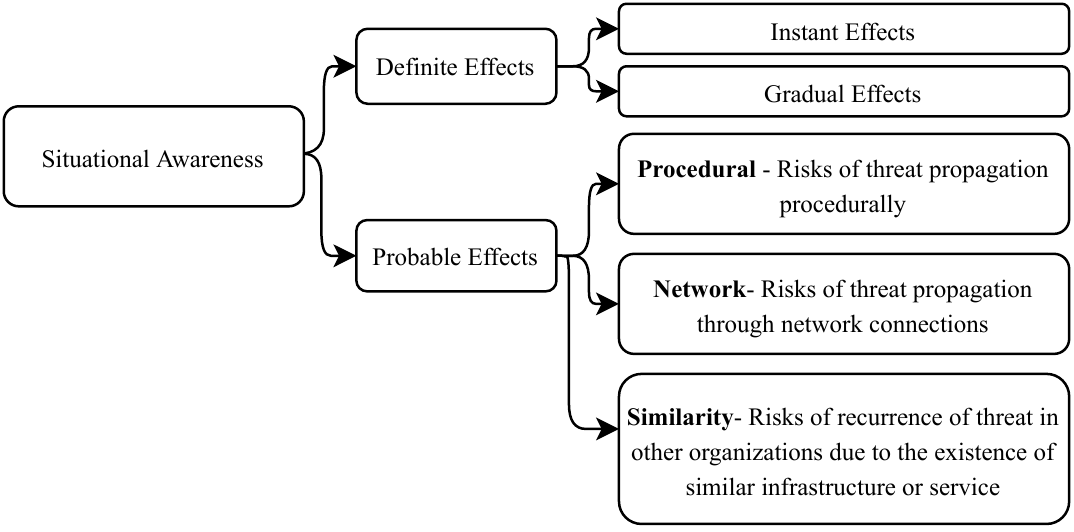}
    \caption{Situational Awareness's detailed components}
    \label{fig:SAcomponent}
\end{figure}

We define two types of SA for a network:
\begin{itemize}
    \item Network's SA for a threat $SA(t_i)$  
    \item Network’s SA for all threats
\end{itemize}
We explain how to calculate the network's SA for a threat $SA(t_i)$ in the following and how to calculate the network's SA for all threats in Section \ref{subsubsec:Network’s SA}.
When a threat occurs, it may cause definite effects at present or probable effects in the near future. Hence, $SA(t_i)$ consists of the definite effects of the threat $t_i$ and projection of its probable effects in the near future on the network.
The definite effect is divided into two parts: Instant definite effect which refers to the effect that a threat has definitely on a service, and gradual definite effect which is the propagation of the threat’s definite effect on the service-organizations depending on that service. 
As mentioned earlier in Section \ref{subsubsubsec:Threat propagation through network connections} threats can propagate on other service-organizations because of similarities in infrastructure or the existence of procedural communications or network connections among the service-organizations. We call risks that may occur as a result of propagation a threat through these three categories probable effects. Therefore, we divide the probable effects into the following three categories:
\begin{itemize}
    \item Risks of threat propagation procedurally 
    \item Risks of threat propagation through network connections
    \item Risks of recurrence of a threat in other organizations due to similar infrastructure or services 
\end{itemize}
Hence, the threat’s SA is calculated and projected according to the Figure ~\ref{fig:SAcomponent}.
\subsection{Definite Effect Calculation}
When a threat of type of attack or incident occurs on a service-organization, the threat may have some definite effects on it. Suppose the other service-organizations in the network depend on and receive a service from that service-organization. In that case, if the threat is an incident, these service-organizations are definitely affected by the threat. Other service-organizations that depend on the second service-organizations and the like are definitely affected by the threat. In this way, this definite effect can be disseminated as a cascade in the network which we call it cascade propagation.
\begin{definition}[Cascade Propagation] \label{def:Cascade Propagation}
Propagation of threat’s definite effect on other service-organizations because of dependencies among them is cascade propagation.
\end{definition}
\begin{example} \label{ex:Example1}
If in Figure ~\ref{fig:Fig1} threat $t_i$ occurs on the service-organization $S_4-O_4$, the service will be affected in proportion to the threat effects which we name it instant definite effect. But due to the cascade propagation, the service-organizations $S_5-O_5$ may also be affected, and again the service-organization $S_6-O_6$, which depends on the service $S_5-O_5$ may also be affected which we call it gradual definite effect. 
\end{example}
As a result, the definite effect of a threat affects the service-organization in which the threat occurs and the network to which the service-organization belongs as shown by red color in Figure~\ref{fig:Fig4}.
The basic formula is used to calculate the definite effect is Equation~\ref{Eq:Eq2}: 
\begin{equation}\label{Eq:Eq2}
Definite\_effect\_of\_the\_threat = Service\_value \times Definitive\_consequences 
\end{equation}
We present Equations~\ref{Eq:Eq3},~\ref{Eq:Eq4} to calculate the instant and gradual definite effect of a threat from the Equation~\ref{Eq:Eq2}.
\begin{align}\label{Eq:Eq3}
    Definite\_Effect(t_i)&= W_{jj} * Crit(Service_j) * CIAAI_j+Imp(S_j)
\end{align}
\begin{align}\label{Eq:Eq4}
    Imp(S_j)&= \sum\nolimits_{k=1}^{m} AW_{jk}*Crit(Service_k)+Imp(S_k)
\end{align}
Suppose the threat $t_i$ occurs on the service-organization $S_j-O_e$. Services that are dependent on the $S_j-O_e$ are $k$ services which $k=1$ to $m$.
\begin{itemize}
    \item $W_{jj}$ is the dependency weight of an service-organization with itself and is always equal to one. 
\item $Imp(S_j)$ is the amount of damage caused by cascading propagation and equal to the gradual definite effect of the threat on dependent service-organizations which is calculated recursively.
\item $AW_{jk}$ is the affected dependency weight between service $j$ and other $k$ dependent services. 
\end{itemize}

To calculate definite effects, we propose an algorithm which Algorithm \ref{algorithm:DECGC} shows the pseudo-code of our algorithm in detail.
By starting from a threatened service, our goal is to find all affected services and calculate instant and gradual definite effects.  
\begin{algorithm}\label{algorithm:DECGC}
	\caption{Definite Effect Calculation and Graph Construction} \label{algorithm:DECGC}
	\renewcommand{\algorithmicrequire}{\textbf{Input:}}
	\renewcommand{\algorithmicensure}{\textbf{Output:}}
	\begin{algorithmic}[1]
		\REQUIRE
		\item[]
			A service-organization dependency graph $\mathcal{G}=(S,D)$ \\
			Type(t), Confimp(t), Intgimp(t), Avlimp(t)\\
			Sid(s), Oid(o)\\
		\ENSURE
		\item[]
			Definite Effect Graph $\mathcal{DE}=(V,A)$ \\
			Definite\_Effect\\ 
		\vspace{0.2cm} 
		\STATE Mark all edges in $D(\mathcal{G})$ as unvisited
		\STATE Set $V(\mathcal{DE})$ and $A(\mathcal{DE})$ to the empty 
		\IF{threat type is equal to Incident or Attack}
		\STATE Calculate definite effect for the first service Sid(s)
		\ENDIF
		\IF {threat type is equal to Incident}
		    \STATE Set visited edges to the empty list and call it $VE$
		    \STATE Add $sid(s)$ to an empty queue $\mathcal{Q}$
		\WHILE {$\mathcal{Q}$ is not empty}
		    \STATE Remove the first element of $\mathcal{Q}$ and call it $r$
		    \STATE Add $r$ to $V(\mathcal{DE})$
		    \STATE Let $\Gamma(r)$ be the neighbor set of $r$ in $\mathcal{D}$
		    \FOR{each node $z\in\Gamma(r)$}
		        \STATE Let $e$ be the edge $(r,z)\in D(\mathcal{G})$
		        \FOR{each entry in $VE$}
		            \IF{e is unvisited}
		                \STATE Add $z$ to the end of $\mathcal{Q}$
                        \STATE Add $e$ to $VE$
		                \STATE Draw an edge from $r$ to $z$ in $\mathcal{DE}$
		                \STATE add definite effect $(r,z)$ to definite effect
		            \ENDIF
		        \ENDFOR
		    \ENDFOR
		\ENDWHILE
		\STATE Add definite effect to the first service's definite effect
    	\STATE return definite effect
        \ENDIF
        \STATE return first service's definite effect 
        \end{algorithmic}
\end{algorithm}
\subsection{Probable Effect}
As shown in Figure \ref{fig:SAcomponent}, the probable effect is divided into three categories. We defined these three categories in Section \ref{subsec:Threat Propagation}. In the following, these three categories and how we calculate them are presented. 
we use Equation~\ref{Eq:Eq7}~\cite{boehm1989software} to calculate these risks.
\begin{equation}\label{Eq:Eq7}
Probable\_effect\_of\_a\_threat = Impact \times Possibility\_of\_propagation \times Service\_value
\end{equation}
\subsubsection{Risks of threat propagation procedurally}
This category measures the risks that may be posed by the spread of a threat by an organization to other procedurally connected organizations in the network by observing a threat of type of attack or incident. We call this part of SA, Procedural Effect and calculate it through using Equation~\ref{Eq:Eq8}.
\begin{equation}\label{Eq:Eq8}
Procedural\_Effect(t_{i})= P_{A} \sum\nolimits_{k=1}^{n} ((1- P_{E_{k}}) \times P_{Propagation\_Procedure_{ek}} \times \\ Crit (Organization_{k}) * CIAI_{k})
\end{equation}
To calculate procedural effects, we propose an algorithm which Algorithm  \ref{algorithm:PECGC} shows the pseudo-code of our algorithm in detail.
By starting from a service on which a threat has occurred, our goal is to find all affected organizations and calculate procedural effects.
\begin{algorithm}
	\caption{Procedural Effect Calculation and Graph Construction} \label{algorithm:PECGC}
	\renewcommand{\algorithmicrequire}{\textbf{Input:}}
	\renewcommand{\algorithmicensure}{\textbf{Output:}}
	\begin{algorithmic}[1]
		\REQUIRE
		\item[]
			Type(t), Confimp(t), Intgimp(t), Avlimp(t)\\
			$P_A(t)$, Sid(s), Oid(o)\\
		\ENSURE
		\item[]
			Procedural Effect Graph $\mathcal{PE}=(V,A)$ \\
			Procedural\_Effect\\ 
		\vspace{0.2cm} 
		\STATE Set $V(\mathcal{PE})$ and $A(\mathcal{PE})$ to the empty 
		\IF {threat type is equal to Incident or Attack}
		    \STATE Add $o$ to  $V(\mathcal{PE})$
		    \STATE Add organizations that has procedural relationship to the Oid(o) to a list and call it $PRL$
		    \FOR{each node $y\in PRL$}
		        \STATE Draw an edge from $o$ to $y$ in $PE$
		        \STATE  add procedural effect $(o,y)$ to procedural effect 
		    \ENDFOR
		    \STATE return procedural effect
        \ENDIF
        \end{algorithmic}
\end{algorithm}
\subsubsection{Risks of threat propagation through network connections}\label{subsubsec:neteff}
This category projects the risks of spreading the threat through network connections with other organizations by observing a threat of type of attack or incident. We name this part of SA Network Effect and calculate it through Equation~\ref{Eq:Eq9}.
\begin{equation}\label{Eq:Eq9}
\begin{split}
NetworkE\_ffect(t_{i})= P_{A} \sum\nolimits_{k=1}^{r} ((1- P_{E_{k}}) \times P_{Propagation\_Network_{ek}} \times Crit (Organizatio_{k}) * CIAI_{k})
\end{split}
\end{equation}
$P_{{Propagation-Network_{jl}}} $ is the probability of spreading the threat through the network connection between two organizations. To calculate this probability, we consider the privileges that an attacker obtains through the threat because she/he can extend her/his threat to other connected organizations through the network based on the obtained privileges. To determine these probabilities, we use Snort's~\cite{snort:2021} attack classifications as shown in Table \ref{tbl:Table4}.
\subsubsection{Risks of recurrence of a threat in other organizations due to similar infrastructure or service}\label{sebsubsec:infraeff}
In this category, risks of recurrence of a threat in other organizations due to similar infrastructure or services are projected by observing all kinds of threats. For example, when a threat occurs on Windows 7 in an organization, this category examines what organizations have Windows 7, how likely they are threatened, how malicious the threat may be, if it occurs, and calculates the extent of its future damage. We name this part of SA Infrastructural Effect and calculate it through using Equation~\ref{Eq:Eq10}.
\begin{equation}\label{Eq:Eq10}
\begin{split}
Infrastructural\_Effect_{t_{i}}= {P}_{A} \sum\nolimits_{k=1}^{q}((1-{P}_{E_ {y}}) \times Crit(Organization_{y}) \times CIAI_{y})
\end{split}
\end{equation}
Since algorithms of Risks of threat propagation through network connections and Risks of recurrence of a threat in other organizations due to similar infrastructure or service are similar to ~\ref{algorithm:PECGC}, we do not mention them. 
Since these four amounts represent $SA(t_i)$ from different dimensions, Network's SA for a threat $SA(t_i)$ is represented as the following quadratic vector:
\begin{equation}\label{Eq:Eq11}
\begin{split}
    SA(t_i)=[Definite\_Effect, Procedural\_Effect, Network\_Effect,  Infrastructural\_Effect]
\end{split}
\end{equation}
We obtain diverse information from each part of this vector. 
\subsubsection{Network’s SA}\label{subsubsec:Network’s SA} 
We presented the method of calculating and projecting the network's SA for a threat $SA(t_i)$ in the form of a quadratic vector so far. To obtain the network's SA for all threats, we add each part of the quadratic vectors which are calculated for each reported threat together. The final vector is also a quadratic vector. Equation~\ref{Eq:Eq12}, shows how to calculate a network's SA.
\begin{equation}\label{Eq:Eq12}
\begin{split}
SA(Network)= SA(Network) + SA(t_i) 
\end{split}
\end{equation}
\subsubsection{SA Reduction}
Over time the observed threats' effects should be reduced by doing and making the necessary actions and decisions For this purpose, by receiving feedback from the organization that has reported threat $t_i$, the amount of $SA(t_i)$  is reduced from the SA(Network) through Equation~\ref{Eq:Eq13}.         
\begin{equation}\label{Eq:Eq13}
\begin{split}
SA(Network)= SA(Network) - SA(t_i) 
\end{split}
\end{equation}
\section{Time Complexity Analysis}
In this section, we analyze the worst-case time complexity of the main algorithms of QR-SACP, namely Definite Effect Calculation and Graph Construction (Algorithm~\ref{algorithm:DECGC}) and Procedural Effect Calculation and Graph Construction (Algorithm~\ref{algorithm:PECGC}).
\subsection{Definite Effect Calculation and Graph Construction}
In this algorithm, we first mark all edges of the service dependency graph as unvisited. Then by starting from an initial service, we find all resources that have forward dependency relationships with that initial service. To track forward dependency relationships, we first construct an empty service-organization dependency graph and then add the initial service to it. For all the services that are direct or indirect dependent on the service (the queue $\mathcal{Q}$) where $n$ is the total number of them in the worst case which is equal to $S(\mathcal{G})$, then for all neighbors of a service where $n$ is the maximum number of a service's neighbors in the worst case and is equal to $S(\mathcal{G})$, if edge between that service and its neighbor is unvisited, we calculate the amount of its definite effect and add it to the total definite effect, which takes $O(n^2)\times O(unvisited-check)$ time in the worst case. Task unvisited-check takes $O(n^2)$ time in the worst case, where $n$ is the total number of services. Therefore, we conclude that the worst-case time complexity of the definite effect calculation and graph construction is $O(n^4)$.
\subsection{Procedural Effect Calculation and Graph Construction}
In this algorithm, we extract organizations that have a procedural relationship with the under threat organization. This task takes $O(r)$ time, where $r$ is the total number of organizations. Then for all organizations that have a procedural relationship with the under threat organization, we calculate procedural effect which takes $O(r)$ time. Therefore, we conclude that the worst-case time complexity of the procedural effect calculation and graph construction is $O(r)$.

\section{Evaluation}\label{sec:result}
In this section, we evaluate the effectiveness of QR-SACP for calculating and projecting situational awareness. 
\subsection{Evaluation Lab}
Threats data that have been used during the evaluation are real. We used Information and Communication Technologies-Information Sharing and Analysis Center (ICT-ISAC) threat data to evaluate QR-SACP. ICT-ISAC receives threat data from ICT member organizations and shares them with other ICT member organizations to increase the security of ICT sector. Since the values used for service, organization, and threat specifications and how to extract them are essential for evaluation, we used aliases to anonymize to preserve their privacy. Among the services that send threat information to ICT-ISAC are Internet Service providers, VoIP service providers, telecommunications infrastructure and hosting service providers. Whether the service name is ISP or $S_{11}$ does not make a difference in evaluating the effectiveness of the  proposed algorithms. 

\begin{figure*}
    \centering
    \includegraphics[width=0.6\textwidth]{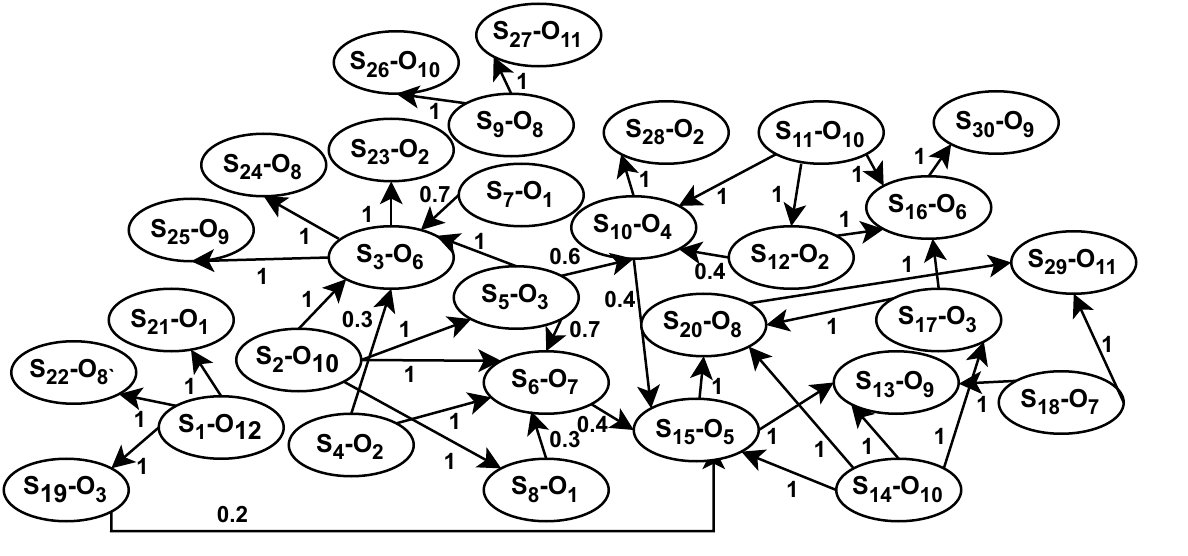}
    \caption{The Evaluation Network}
    \label{fig:Fig3}
\end{figure*}

In this evaluation, threat data belong to 30 critical services from 12 ICT-ISAC member organizations. 
Tables \ref{tbl:orgprop}, \ref{tbl:srvprop}, ~\ref{tbl:procthpro}, ~\ref{tbl:Table4} and ~\ref{tbl:sumthre} shows organization information, service information, probability of threat propagation procedurally, probability of threat propagation through network and summary of threats and the obtained results, respectively. The dependencies among services of organizations has been shown in Figure ~\ref{fig:Fig3}. 

Questionnaires were presented to the organizations to receive their assets $CPE_{id}$ and the values of service and organization CIA Demand, ${P_e}$, procedural probabilty propagation between organizations, $P_{net}$. The legislator finalized the received values based on the importance of the services and organizations in the ICT sector.
Criticality of services and organizations has been set by the legislator.
Threat CIA has been collected from ~\cite{nvd:2021}.
\begin{table}[!t]
\begin{minipage}[c]{0.5\textwidth}
\caption{Service Information}
	\label{tbl:srvprop}
	\begin{tabular}{cp{0.6cm}ccc}\hline
		\bfseries Service & \bfseries $O_{id}$ & \bfseries Criticality & \bfseries CIA Demand & \bfseries $\boldsymbol{P_e}$ \\\hline
		$S_1$ &	$O_{12}$ & 0.8 & (0.6, 0.2, 0.9) & 0.8 \\
		$S_2$ &	$O_{10}$ & 0.9 & (0.1, 0.2, 0.9) & 0.1 \\
		$S_3$ &	$O_6$ & 0.7 & (0.9, 0.6, 0.8) & 0.4 \\
		$S_4$ &	$O_2$ & 0.6 & (0.4, 0.1, 0.7) & 0.3 \\
		$S_5$ &	$O_3$ & 0.3 & (0.2, 0.1, 0.7) & 0.3 \\
		$S_6$ &	$O_7$ & 0.6 & (0.5, 0.5, 0.7) & 0.7 \\
		$S_7$ &	$O_1$ & 0.5 & (0.4, 0.3, 0.7) & 0.4 \\
		$S_8$ &	$O_1$ & 0.6 & (0.9, 0.2, 0.7) & 0.4 \\
		$S_9$ &	$O_8$ & 0.3 & (0.3, 0.6, 0.8) & 0.2 \\
		$S_{10}$ &	$O_4$ & 0.5 & (0.4, 0.5, 0.7) & 0.3 \\
		$S_{11}$ &	$O_{10}$ & 0.8 & (0.2, 0.1, 1) & 0.9 \\
		$S_{12}$ &	$O_2$ & 0.5 & (0.2, 0.3, 0.6) & 0.5 \\
		$S_{13}$ &	$O_9$ & 0.4 & (0.2, 0.3, 0.7) & 0.4 \\
		$S_{14}$ &	$O_{10}$ & 0.8 & (0.2, 0.3, 1) & 0.6 \\
		$S_{15}$ &	$O_5$ & 0.7 & (0.2, 0.6, 0.8) & 0.5 \\
		$S_{16}$ &	$O_6$ & 0.1 & (0.6, 0.4, 0.7) & 0.3 \\
		$S_{17}$ &	$O_3$ & 0.3 & (0.3, 0.1, 0.7) & 0.2 \\
		$S_{18}$ &	$O_7$ & 0.1 & (0.5, 0.2, 0.6) & 0.3 \\
		$S_{19}$ &	$O_3$ & 0.5 & (0.2, 0.2, 0.6) & 0.6 \\
		$S_{20}$ &	$O_8$ & 0.6 & (0.8, 0.6, 0.2) & 0.5 \\
		$S_{21}$ &	$O_1$ & 0.2 & (0.3, 0.5, 0.8) & 0.3 \\
		$S_{22}$ &	$O_8$ & 0.4 & (0.4, 0.8, 0.3) & 0.3 \\
		$S_{23}$ &	$O_2$ & 0.1 & (0.2, 0.1, 0.3) & 0.2 \\
		$S_{24}$ &	$O_8$ & 0.3 & (0.2, 0.4, 0.5) & 0.1 \\
		$S_{25}$ &	$O_9$ & 0.3 & (0.3, 0.6, 0.3) & 0.4 \\
		$S_{26}$ &	$O_4$ & 0.1 & (0.2, 0.3, 0.6) & 0.1 \\
		$S_{27}$ &	$O_{11}$ & 0.2 & (0.3, 0.5, 0.2) & 0.3 \\
		$S_{28}$ &	$O_2$ & 0.4 & (0.3, 0.6, 0.2) & 0.2 \\
		$S_{29}$ &	$O_{11}$ & 0.2 & (0.1, 0.1, 0.1) & 0.3 \\
		$S_{30}$ &	$O_9$ & 0.2 & (0.4, 0.2, 0.1) & 0.5 \\\hline
	\end{tabular}
	\end{minipage}
\begin{minipage}[c]{0.5\textwidth}
	\caption{Network Probability Propagation}
	\label{tbl:Table4}
	\centering
	\begin{tabular}{lc}\hline
		\bfseries Classification & \bfseries Prob.  \\\hline
	    Successful-admin &	1 \\
		Trojan-activity &	1 \\
		Shellcode-detect &	1 \\
        Web-application-attack &	0.9 \\
		Unauthorized access to data &  0.9\\
		Successful-user &	0.85 \\
		Successful-recon-largescale & 0.7 \\
		Denial-of-service &	 0.5\\
        Attempted-admin &	0.4 \\
		Attempted-user &	0.3 \\
		Default-login-attempt &	0.3 \\
		Suspicious-filename-detect & 0.3\\
    	Suspicious-login & 0.3 \\
    	Scan &  0.2\\
    	Other & 0.1 \\\hline
	\end{tabular}
	\end{minipage}
	\end{table}
\begin{table}[t]
\begin{minipage}[c]{0.5\textwidth}
\centering
\caption{Organization Information}
	\label{tbl:orgprop}
\begin{tabular}{cccc}\hline
		\bfseries Oid & \bfseries Criticality & \bfseries CIA Demand & \bfseries $\boldsymbol{P_e}$ \\\hline
		$O_1$ &	0.5 & (0.5, 0.5, 0.8) & 0.4 \\
		$O_2$ &	0.4 & (0.2, 0.2, 0.7) & 0.2 \\
		$O_3$ &	0.6 & (0.4, 0.1, 0.9) & 0.5 \\
		$O_4$ &	0.7 & (0.8, 0.3, 0.7) & 0.5 \\
		$O_5$ &	0.5 & (0.5, 0.9, 0.8) & 0.3 \\
		$O_6$ &	0.6 & (0.7, 0.5, 0.7) & 0.6 \\
		$O_7$ &	0.5 & (0.7, 0.4, 0.7) & 0.6 \\
		$O_8$ &	0.3 & (0.6, 0.8, 0.6) & 0.4 \\
		$O_9$ &	0.2 & (0.2, 0.2, 0.7) & 0.2 \\
		$O_{10}$ & 0.9 & (0.4, 0.4, 0.9) & 0.8 \\
		$O_{11}$ & 0.3 & (0.2, 0.5, 0.8) & 0.3 \\
		$O_{12}$ & 0.8 & (0.4, 0.3, 0.9) & 0.9 \\\hline
	\end{tabular}
\end{minipage}
\begin{minipage}[c]{0.5\textwidth}
\centering
\caption{Procedural Probability Propagation}
	\label{tbl:procthpro}
	\begin{tabular}{cc}\hline
		\bfseries Oid & \bfseries Oid,Probability\\\hline
		$O_1$ &	\{$O_2$,0.2\}, \{$O_6$,0.3\}, \{$O_{10}$,0.5\}  \\
		$O_2$ &	\{$O_1$,0.1\}, \{$O_5$,0.4\}, \{$O_{11}$,0.3\}  \\
		$O_3$ &	\{$O_7$,0.3\} \\
		$O_4$ & - \\
		$O_5$ &	\{$O_2$,0.4\}  \\
		$O_6$ &	\{$O_1$,0.4\}  \\
		$O_7$ &	\{$O_3$,0.6\}  \\
		$O_8$ &	 - \\
		$O_9$ &	 -  \\
		$O_{10}$ & \{$O_1$,0.4\}  \\
		$O_{11}$ & \{$O_2$,0.4\}  \\
		$O_{12}$ & \{$O_8$,0.5\} \\
	\end{tabular}
\end{minipage}
	\end{table}
\begin{table*}[!t]
    \footnotesize
	\caption{Summary of Threats and Results}
	\label{tbl:sumthre}
	\centering
	\begin{tabular}{p{1cm}p{1cm}cccccp{1.7cm}p{1.7cm}p{1.4cm}p{1.65cm}}\hline
		\bfseries Threat Type & \bfseries Service & \bfseries $O_{id}$ & \bfseries Threat CIA & \bfseries $\boldsymbol{P_e}$ & \bfseries $\boldsymbol P_{net}$ & \bfseries $CPE_{id}$ & \bfseries Def. Eff. & \bfseries Proc. Eff. & \bfseries Net. Eff. & \bfseries Infra. Eff.\\\hline
		Inc & $S_{11}$ & $O_{10}$ & (C, C, C) & 1    & --        & 602 & 75.63,75.63 & 0.98,0.98 & 0,0 & 0.47,0.47\\
		Inc & $S_{21}$ & $O_1$    & (C, N, P) & 0.95 & 0.3       & 70  & 0.77,76.4   & 0.98,1.96 & 0.26,0.26 & 0.42,0.89\\
		Inc & $S_{17}$ & $O_3$    & (P, C, C) & 0.9  & {0.3,0.4} & 135 & 25.47,101.87& 0.38,2.34 & 1.15,1.41 & 0.73, 1.62\\
		Inc & $S_3$    & $O_6$    & (C, C, C) & 0.85 & --        & 126 & 25.72,127.59& 0.83,3.17 & 0,1.41 & 0.67,2.29\\
		Inc & $S_4$    & $O_2$    & (P, P, P) & 1    & --        & 439 & 30.16,157.75& 1.08,4.25 & 0,1.41 & 0.11,2.4\\
		Inc & $S_{18}$ & $O_7$    & (P, P, N) & 0.85 & --        & 6   & 4.46,162.21 & 0.21,4.46 & 0,1.41 & 0.89,3.29\\
		Atk & $S_{11}$ & $O_{10}$ & (C, C, C) & 1    & --        & 748 & 6.03,168.24 & 0.98,5.44 & 0,1.41 & 0.11,3.4\\
		Atk & $S_5$    & $O_3$    & (P, C, N) & 1    & {0.3,0.4} & 56  & 0.36,168.6 & 0.25,5.69 & 0.83,2.24 & 0.55,3.95\\
		Atk & $S_7$    & $O_1$    & (P, P, P) & 0.9  & 0.3       & 15  & 1.77,170.37  & 0.8,6.49  & 0.23,2.47 & 0.53,4.48\\
		Atk & $S_2$    & $O_{10}$ & (N, N, N) & 0.95 & --        & 916 & 0,170.37    & 0,6.49    & 0,2.47 & 0,4.48\\
		Atk & $S_{23}$ & $O_2$    & (C, C, P) & 0.85 & --        & 351 & 0.26,170.63 & 1.45,7.94 & 0,2.47 & 0.89,5.37\\
		Atk & $S_1$    & $O_{12}$ & (P, P, N) & 0.9  & 0.5       & 248 & 1.75,172.38 & 0.29,8.23 & 0.18,2.65 & 0.79,6.16\\
		Atk & $S_{20}$ & $O_8$    & (P, P, C) & 1    & 0.2       & 34  & 2.71,175.09 & 0,8.23    & 0.4,3.05 & 0.15,6.31\\
		Atk & $S_6$    & $O_7$    & (N, N, C) & 1    & --        & 166 & 2.88,177.97 & 1.11,9.34 & 0,3.05 & 0.25,6.56\\
		Atk & $S_{12}$ & $O_2$    & (N, N, P) & 1    & --        & 446 & 0.85,178.82 & 0.53,9.87 & 0,3.05 & 0.54,7.1\\
		Atk & $S_8$    & $O_1$    & (C, C, C) & 0.9  & 0.3       & 281 & 5.06,183.88 & 1.55,11.42& 0.41,3.46 & 0.88,7.98\\
		Vul & $S_{11}$ & $O_{10}$ & (P, P, C) & 0.85 & --        & 104 & 0,183.88    & 0,11.42   & 0,3.46 & 0.25,8.23\\
		Vul & $S_3$    & $O_6$    & (C, C, C) & 0.85 & --        & 729 & 0,183.88    & 0,11.42   & 0,3.46 & 0.4,8.63\\
		Vul & $S_{17}$ & $O_3$    & (N, N, N) & 1    & --        & 523 & 0,183.88    & 0,11.42   & 0,3.46 & 0,8.63\\
		Vul & $S_9$    & $O_8$    & (P, P, P) & 0.9  & --        & 281 & 0,183.88    & 0,11.42   & 0,3.46 & 0.65,9.28\\
		Vul & $S_{23}$ & $O_2$    & (P, C, N) & 0.9  & --        & 104 & 0,183.88    & 0,11.42   & 0,3.46 & 0.35,9.63\\
		Vul & $S_{29}$ & $O_{11}$ & (N, N, N) & 1    & --        & 135 & 0,183.88    & 0,11.42   & 0,3.46 & 0,9.63\\
		Vul & $S_{24}$ & $O_8$    & (P, P, P) & 0.95 & --        & 869 & 0,183.88    & 0,11.42   & 0,3.46 & 0.81,10.44\\
		Vul & $S_{7}$  & $O_1$    & (C, C, P) & 0.95 & --        & 149 & 0,183.88    & 0,11.42   & 0,3.46 & 0.38,10.82\\
		Vul & $S_{10}$ & $O_4$    & (C, C, C) & 0.95 & --        & 255 & 0,183.88    & 0,11.42   & 0,3.46 & 0.11,10.93\\
	\end{tabular}
\end{table*}
In section ~\ref{sec:ASSmdl} each of these terms is introduced in detail.
\subsection{Evaluation Results}
We received 25 threats during the 30-day evaluation period through threat information sharing shown in Table ~\ref{tbl:sumthre}. Ten of these threats are vulnerabilities, nine are attacks and six are incidents. 

\begin{figure}
\centering
\begin{minipage}{.5\textwidth}
  \centering
  \includegraphics[width=1\linewidth]{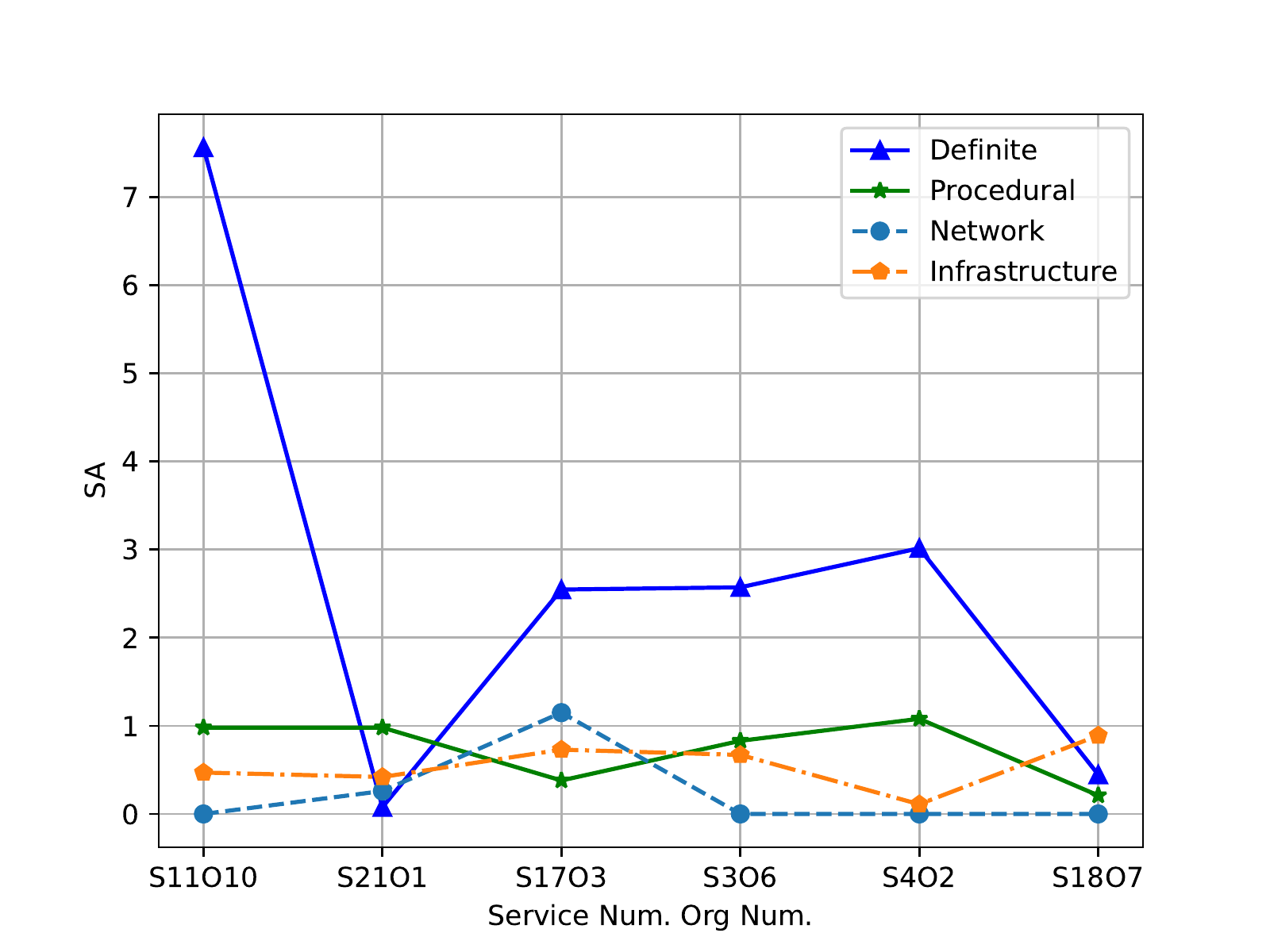}
  \caption{Incidents' SA vector values}
  \label{fig:incitak}
\end{minipage}%
\begin{minipage}{.5\textwidth}
  \centering
  \includegraphics[width=1\linewidth]{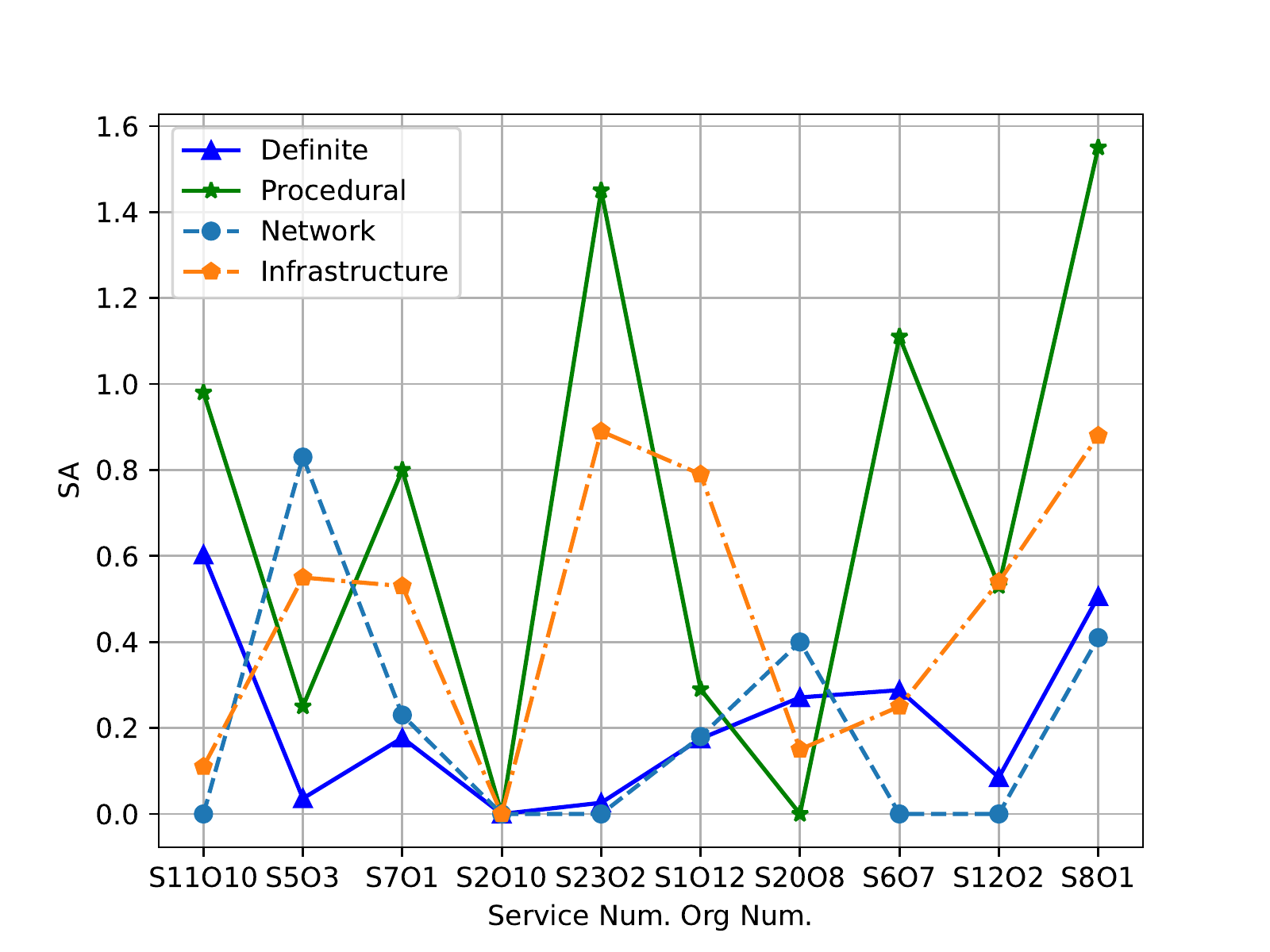}
  \caption{Attacks' SA vector values}
  \label{fig:atktak}
\end{minipage}
\centering
\begin{minipage}{.5\textwidth}
  \centering
  \includegraphics[width=1\linewidth]{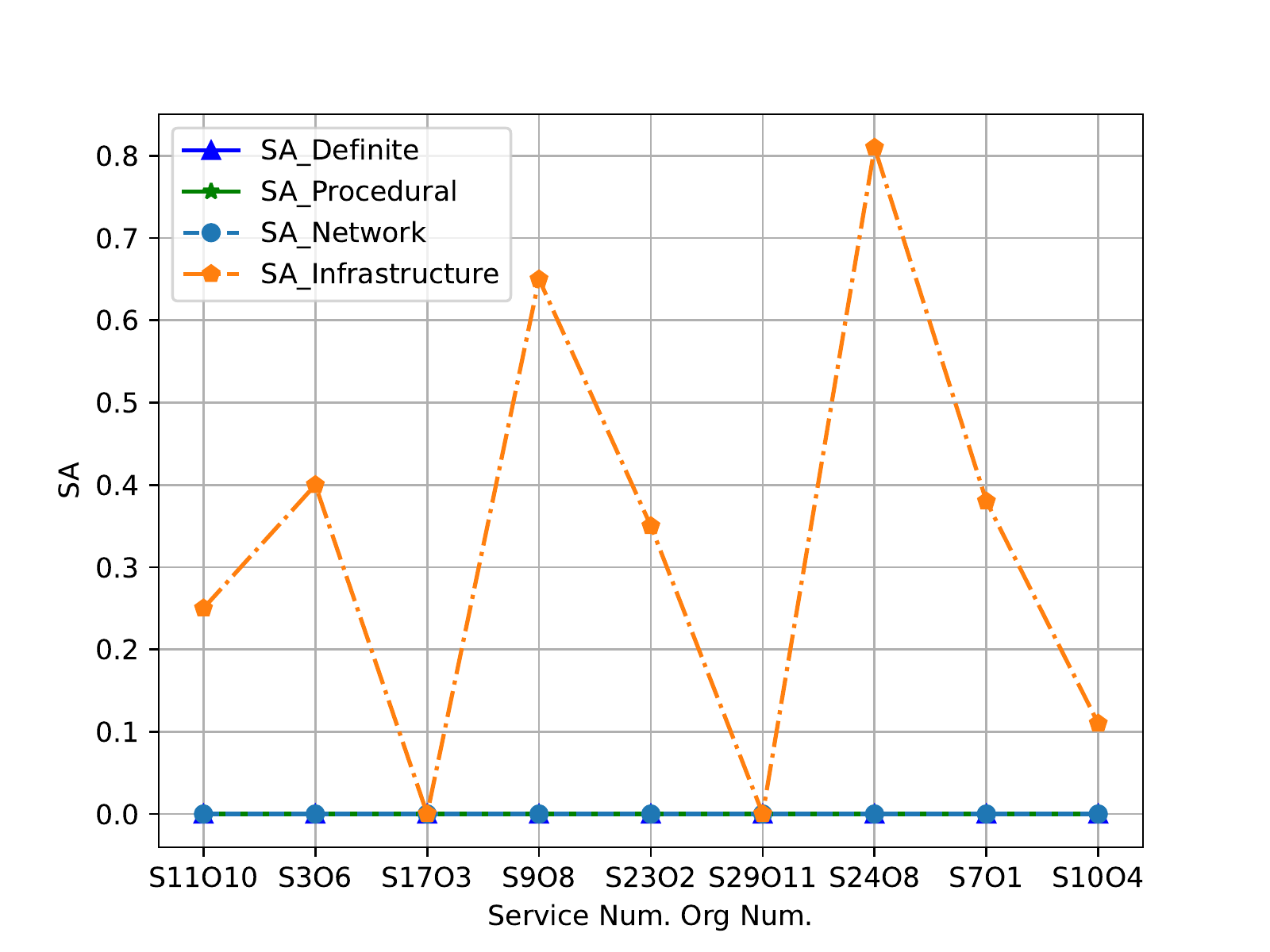}
  \caption{Vulnerabilities' SA vector values}
  \label{fig:vultak}
\end{minipage}%
\begin{minipage}{.5\textwidth}
  \centering
  \includegraphics[width=1\linewidth]{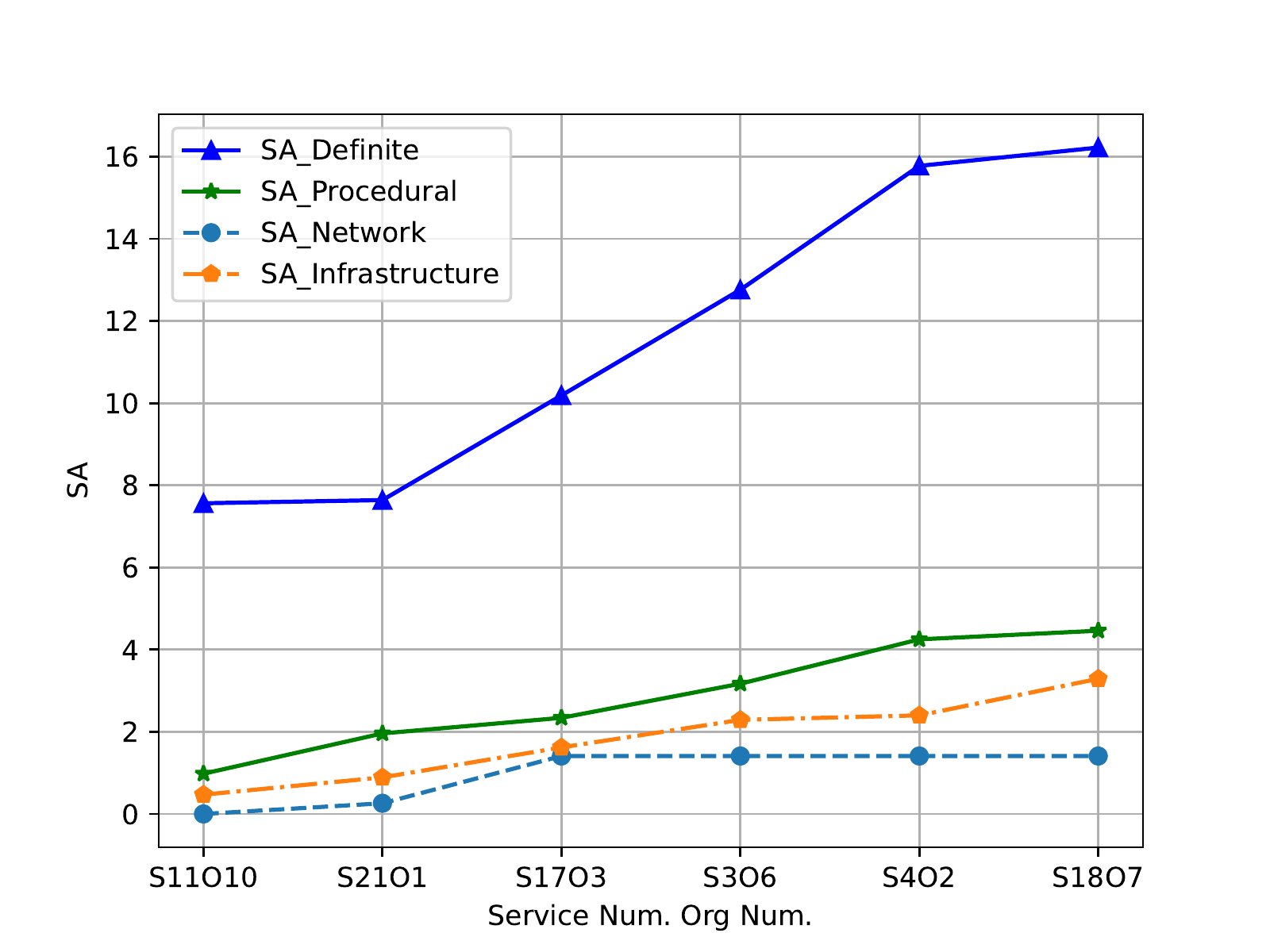}
  \caption{Network's SA vector values 1}
  \label{fig:suminci}
\end{minipage}
\centering
\begin{minipage}{.5\textwidth}
  \centering
  \includegraphics[width=1\linewidth]{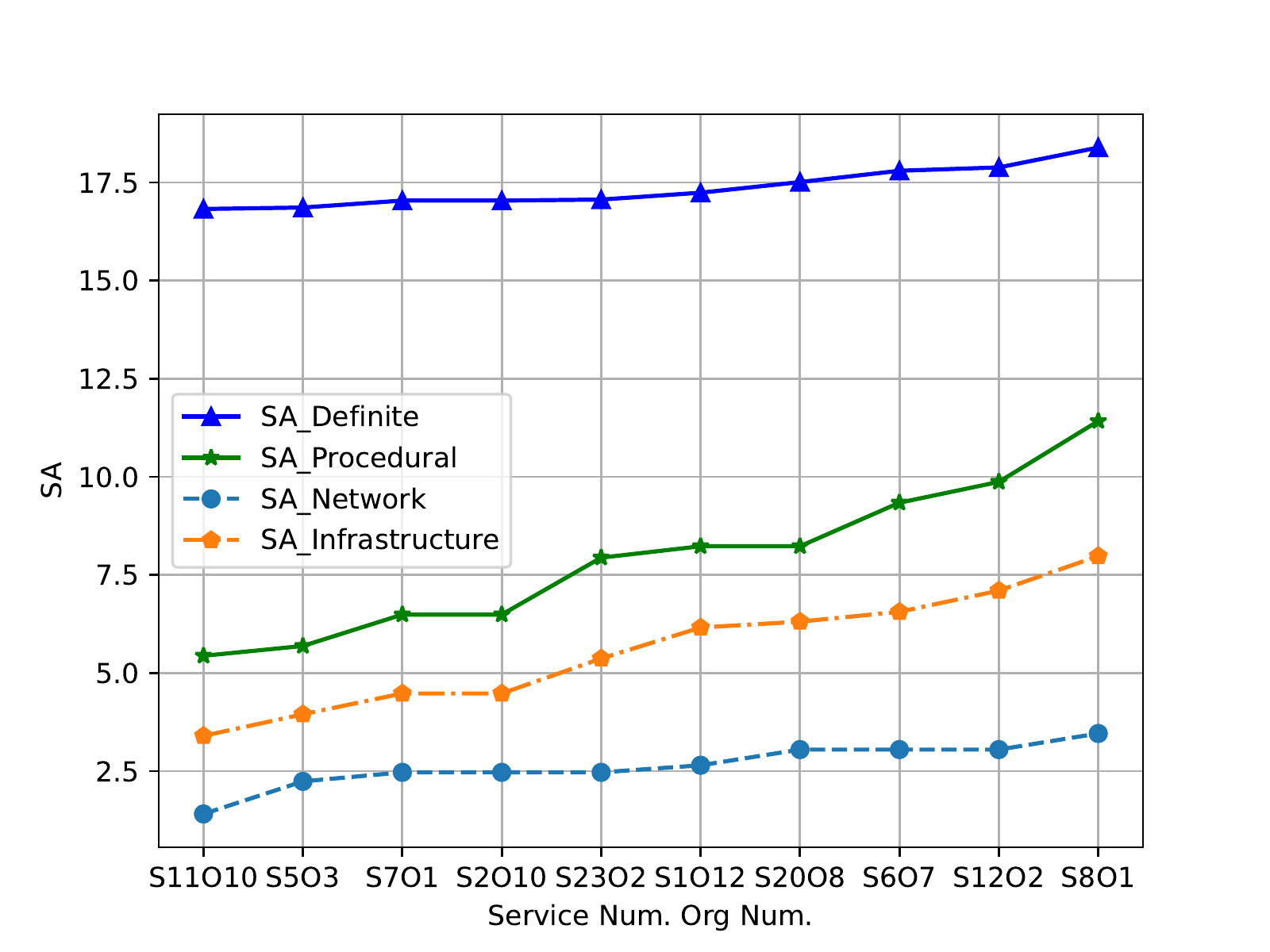}
  \caption{Network's SA vector values 2}
  \label{fig:sumatk}
\end{minipage}%
\begin{minipage}{.5\textwidth}
  \centering
  \includegraphics[width=1\linewidth]{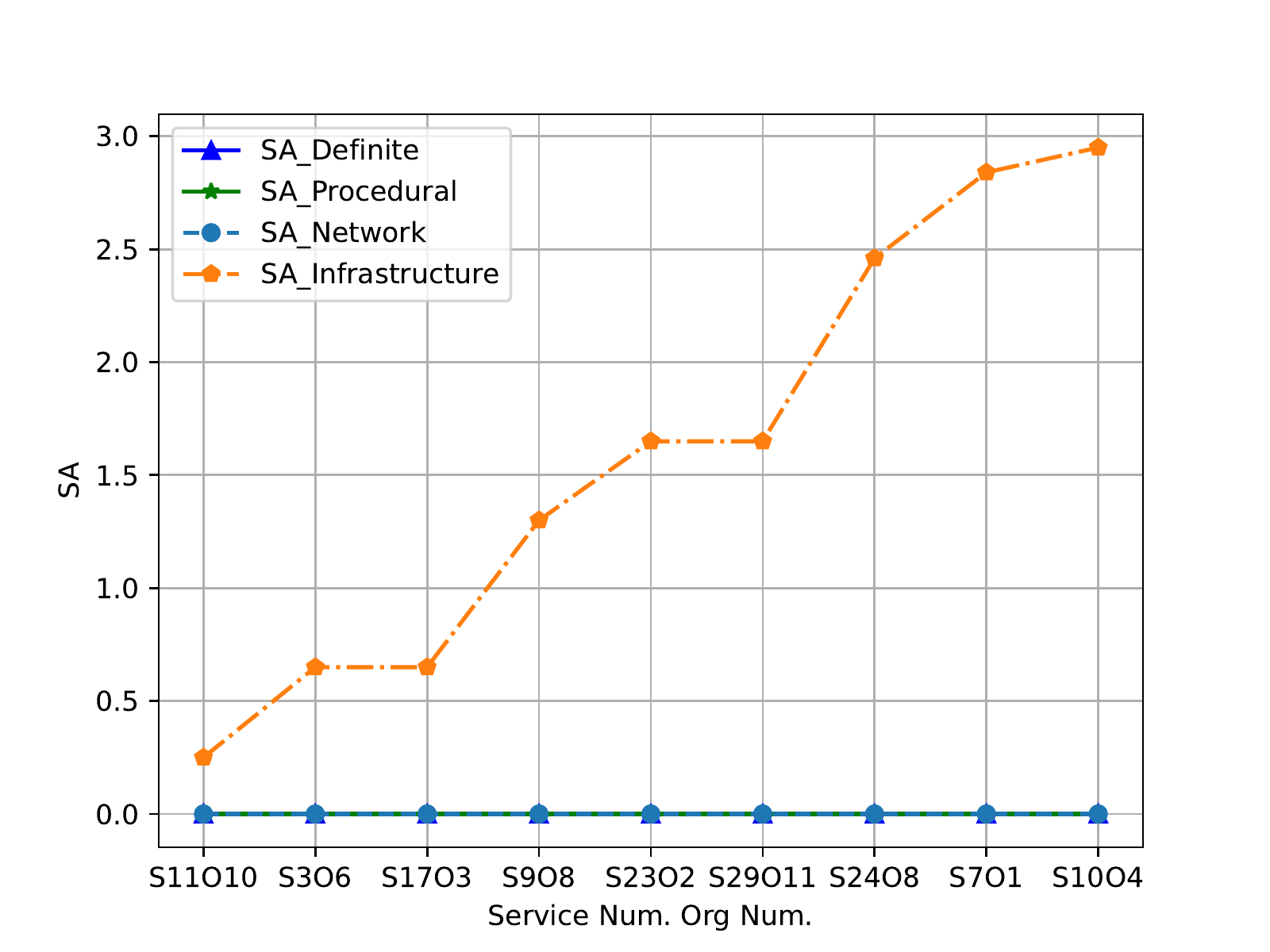}
  \caption{Network's SA vector values 3}
  \label{fig:sumvul}
\end{minipage}
\end{figure}

\begin{figure}
    \centering
    \includegraphics[width=0.6\textwidth]{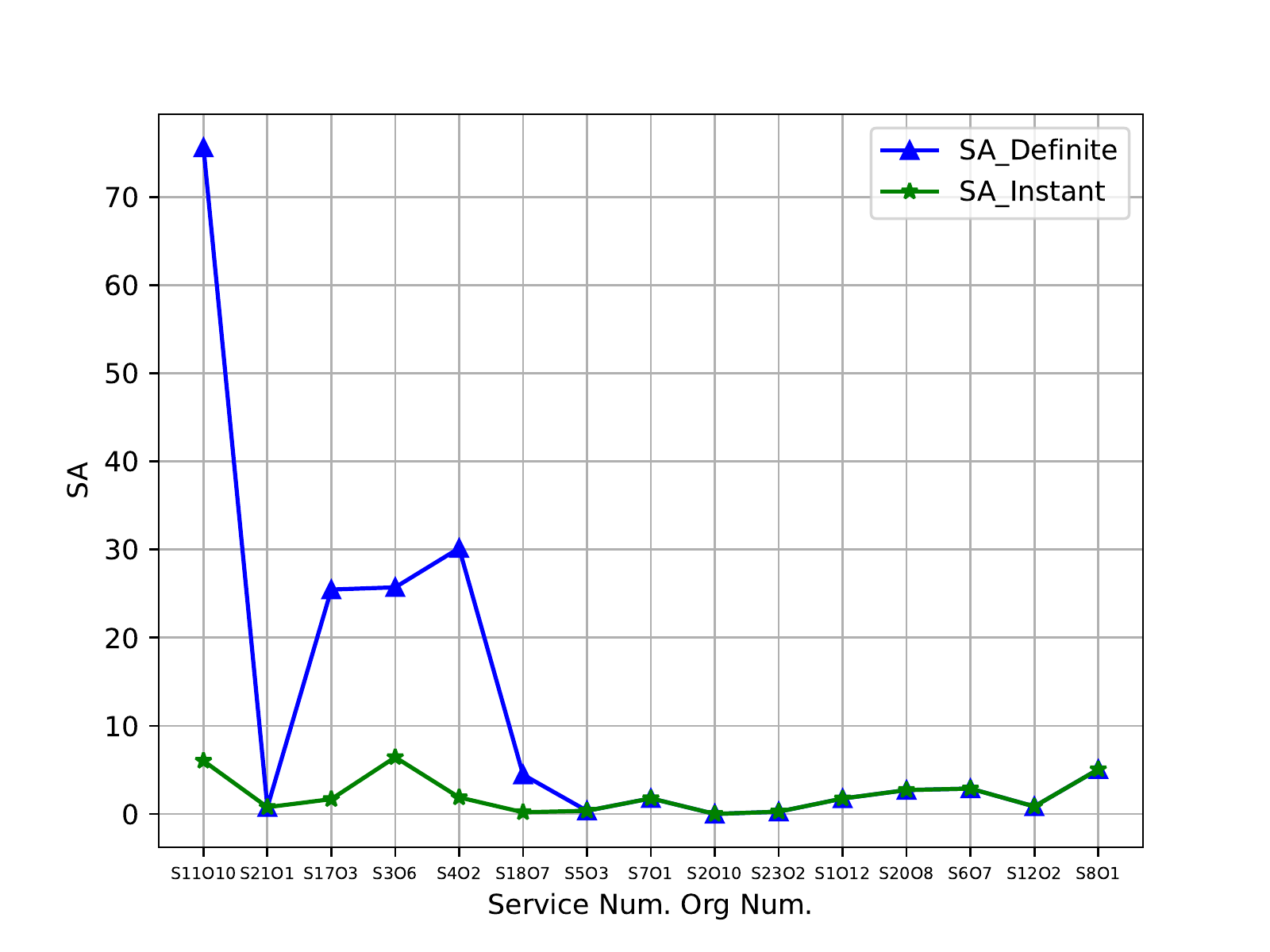}
    \caption{Instant Definite Effects in comparison to Definite Effect }
    \label{fig:insdefi}
\end{figure}

Figure ~\ref{fig:incitak}, ~\ref{fig:atktak} and ~\ref{fig:vultak} shows calculated threat's SA vector values for incidents, attacks and vulnerabilities, respectively.  
Network's SA vector values for incident, attack and vulnerability in Figure ~\ref{fig:suminci}, ~\ref{fig:sumatk} and ~\ref{fig:sumvul} are shown, respectively. In columns 8 to 11 of Table~\ref{tbl:sumthre}, the first value is for each threat, and the second value is the sum of the values for all threats up to that threat. The Network's SA vector values for the type of incident have been divided by ten due to a more appropriate representation in the graph. 
Figure~\ref{fig:insdefi} shows instant definite effect and definite effect values for incidents and attacks. 
This way of calculating and projecting threats' SA allows us to compare threats' effects on a service and across the network together from different aspects, including definite effects on the service and its dependent services, procedural propagation, network propagation and infrastructural propagation. In this way, we can choose the most appropriate and adequate timely decisions in response to the situation that a threat has engendered or may engender in the near future in proportion to types of its effects to reduce costs and consequences. For example, to deal with the incident that has occurred in $S_{11}-O_{10}$ we know we should pay attention to its dependent services in comparison to the attack that has occurred in $S_{23}-O_{2}$ that we should pay attention to its dependent organizations which have procedural relationship with it.
Complementary effect of SA vector's parts are well shown in Figure ~\ref{fig:incitak}, ~\ref{fig:atktak} and ~\ref{fig:vultak}. 
For each threat, we understand the definite effect that the threat has left on the network, the probability that the threat may propagate in other organizations procedurally, the probability that the threat may propagate in other organizations through network connections and the probability that the threat occurs again in other organizations because of the similarity of infrastructure or services. 
Furthermore, in Figure \ref{fig:insdefi} if we just calculated and considered instant definite effect, we could not give more priority to $S_{11}-O_{10}$ in comparison to $S_{3}-O_{6}$.
\subsection{Comparison}
four well-known studies that have been done in calculation SA scope are~\cite{alavizadeh2020cyber},~\cite{Zhang2021Network}, ~\cite{rongrongframework} and ~\cite{kou2019research}. Since the output of QR-SACP and the four mentioned studies are values and whether the numbers are larger or smaller does not represent the superiority of the methods over each other and is not a basis for comparing them, we make a qualitative comparison between QR-SACP and the four mentioned studies. 
The calculation method in ~\cite{alavizadeh2020cyber} considers SA calculation equal to risk calculation by considering just vulnerabilities. Therefore, it cannot calculate and project attacks' and incidents' effects. In addition, ~\cite{alavizadeh2020cyber} does not consider dependencies among services in the network. Therefore, it cannot pay attention to cascade propagation in decision-making. For example, ~\cite{alavizadeh2020cyber} selects threats number $S_{18}-O_{7}$, $S_{23}-O_{2}$ or $S_{8}-O_{1}$ to investigate first and ignores threats number $S_{11}-O-{10}$, $S_{17}-O_{3}$, and $S_{3}-O_{6}$ that we should give them more priority. 
Also, ~\cite{Zhang2021Network}, ~\cite{rongrongframework} and ~\cite{kou2019research} do not consider the possibility of spreading threats in the network procedurally, through a network connection and the similarity among services and infrastructures to project threats effect. In addition, they do not consider dependency among systems in a network. They can calculate values like the values green diagram in Figure ~\ref{fig:insdefi} for incidents or attacks, but it cannot predict that threats $S_{11}-O{10}$, $S_{4}-O{2}$, and $S_{3}-O{6}$ will have further destructive effects in the network in the near future. Moreover, they cannot project what effects threats have on the network infrastructure, like Figure ~\ref{fig:vultak}. Furthermore, ~\cite{rongrongframework} calculates the average of SAs as the network's SA and does not consider the summation of SAs. Hence, we cannot have an integrated perspective from the network situation in this method like what we have in Figure ~\ref{fig:suminci}, ~\ref{fig:sumatk} and ~\ref{fig:sumvul}. What ~\cite{kou2019research} calculates is equal to the summation values definite effects and network effects of QR-SACP. Therefore, it cannot calculate threats' procedural and infrastructural effects. In addition, because ~\cite{kou2019research} sums the two values definite effects and network effects together, it cannot determine whether the threat has been happened or may happen in the future. 

\section{CONCLUSION and FUTURE WORKS}\label{sec:CONCLUSION}
In this paper, we proposed QR-SACP, a novel technique that investigates a threat from different aspects through using diverse resources to calculate and project a network's SA for each and all threats. In this technique, a SA's quadratic vector is calculated and projected by receiving a threat through threat information sharing, in proportion to the type of threat. We investigate the threat from different dimensions and extract information, including the threat's properties, infrastructure and configuration on which the threat occurs, impacts that the threat has and ways the threat can propagate across the network and contaminate other services and organizations. If the threat is an incident or attack, all parts of the SA's vector are calculated otherwise only the fourth part of the SA's vector is calculated. We calculate the instant definite effect of the threat on a service. If the threat is an incident, it may have definite effects on the service, such as service interruptions and breakdowns, loss of data confidentiality, and so on. Since other services in the network may depend on the threatened service, the threat may affect these dependent services, too. Therefore, we calculate the propagation of the threat's definite effects across the network and name it gradual definite effect. To calculate gradual definite effect, we consider service dependencies in the network and model them by a weighted directed graph called a service dependency graph. Furthermore, by calculating procedural effects, network effects and infrastructural effects, we project probability of propagation or recurrence of the threat in other network's services and organizations through three categories, namely threat propagation procedurally, threat propagation through network connections, recurrence of a threat in other organizations due to similar infrastructure or services. The experimental results demonstrate QR-SACP method can calculate and project definite and probable threat's effects across the entire network and reveals more details from the threat's current and near future situation to make timely decisions and reduce threat's costs and consequences.

\bibliographystyle{splncs04}

\bibliography{ref.bib}

\end{document}